# Asset Administration Shell-Based OCL Validation Framework for Model-Based System Engineering

Om Parkash* Jannik Bauer* Vincent Schmitt* Thomas Greiner* Rainer Drath*

*IOS³ – Institute of Smart Systems and Services, Pforzheim University of Applied Sciences, Pforzheim, Germany
(e-mails: {om.parkash, jannik.bauer, schmittv, thomas.greiner, rainer.drath}@hs-pforzheim.de)

**Abstract**: Increasing complexity of modern enterprise systems and the demand for automation and interoperability require consistent and semantically validated models in Model-Based Systems Engineering (MBSE). The Object Constraint Language (OCL) supports formal definition of such constraint validations. However, MBSE models and OCL constraints are typically managed in separate tools, causing manual effort during model constraint application and result interpretation. To address this gap, this paper proposes an approach to managing OCL constraints and their validation results through Asset Administration Shells (a well-established technology for interoperability in enterprise systems). The methodology is demonstrated through a fictional industrial scenario, and to support reproducibility, all artifacts are publicly available in a GitHub repository[1].

*Keywords*: MBSE, OCL, AAS, Semantic Constraint Modeling, AutomationML

## 1. INTRODUCTION

Model-Based Systems Engineering (MBSE) is a formal approach to system development that relies on modeling (Bahill, 2018) (International Council on Systems Engineering (INCOSE), 2007). MBSE methods are increasingly being adopted across industries because of their advantages over traditional document-based approaches (Boggero et al., 2021). These advantages include improved system design quality, better traceability of requirements, and clearer communication among development teams (Boggero et al., 2021) (Lemazurier et al., 2017).

Different engineering domains have successfully applied MBSE to develop complex systems. For example, (Oudina et al., 2023). presents the design of a modern industrial control and SCADA system for the oil and gas industry using an MBSE approach. (Castaing et al., 2023) provides a detailed example of applying MBSE to a planetary rover design case study. (Zheng et al., 2024) demonstrates the use of MBSE in constructing multi-energy coupling systems to manage adjustable loads and improve energy efficiency. (Bergelin et al., 2022) proposes a MBSE methodology for evaluating the feasibility of heterogeneous battery systems in the railway domain.

In MBSE, the faithfulness, correctness and maintainability of models are central to developing robust systems, enabling analysis, verifying behavior, and supporting system evolution across different lifecycle phases (Mohagheghi & Aagedal, 2007). In Model-Driven Engineering (MDE), the Object Constraint Language (OCL) is used; it is a textual specification language to define and enforce semantic constraints on data models. It provides a formal way to specify rules as constraints to ensure the correctness of model elements (Gogolla, 2013).

However, OCL constraints and MBSE models are often modelled, stored and managed in separate models, tools or repositories. Typically, MBSE models are loaded into an environment that supports OCL, where the constraints are executed on metamodel instances. Once the constraint validation results are available, they are usually transferred to other systems manually or through scripts for further processing. As MBSE approaches are increasingly applied in industrial ecosystems, a challenge arises in organizing such semantic constraints in a systematic, scalable, and interoperable way.

The Asset Administration Shell (AAS) is already a well-known technology that supports interoperability, transparency, and efficient data usage in Industry 4.0 and is currently establishing itself in industry as a digital representation of assets, i.e. components (OPC Foundation, 2025). AAS is therefore the first methodical and systematic approach to describing industrial systems at the component level in a standardized manner across manufacturers using semantic data models. To do this, standardized submodels are combined. However, even at this level, there is no possibility of defining OCL constraints in AAS or managing their validation results using AAS itself. This contribution therefore proposes a novel approach of combining constraints with AAS and suggests a complete workflow for managing semantic OCL constraints for information models using AAS by creating the necessary submodels. In this approach, AAS acts as a central integration layer that connects models and constraint-related data. The authors present a methodology along with the required components to achieve this. In the end, a fictional industrial scenario is created which demonstrates the workflow, and all related files and scripts are available in a public GitHub repository[1]. The rest of the paper is organized as follows:

---

[1] GitHub Repository: *https://github.com/omparkash-unipf/MBSE_AAS_OCL/tree/main*

Chapter II reviews related work, Chapter III explains the core idea and methodology, Chapter IV presents the methdodology implementation workflow, Chapter V presents the application with a use case example and Chapter VI summarizes the main results and outlines future work.

## 2. RELATED WORK

OCL (Object Management Group (OMG), 2014) (Cabot & Gogolla, 2012) is a textual language for specifying semantic constraints and has been applied in various domains. For example, (Siala & Lano, 2025) uses large language models to automatically derive OCL constraints from existing Java/Python code. (Ahmed et al., 2024) employs OCL as a formal specification language to define ATM system behavior and detect errors early. Additionally, (Ali et al., 2014) reports practical guidance and industrial experiences in applying OCL across different projects. A study done by (Ferdjoukh et al., 2015) focuses on embedding OCL constraints into model generation using constraint programming.

Related work has also examined the storage of OCL validation results in files. (Barisas & Bareisa, 2009) demonstrate storage of OCL-based artifacts—such as generated test paths—as well as detailed execution results, including object states, in dedicated files. Similarly (Izsó et al., 2014) propose a framework for assessing the scalability of MDE approaches by validating models using hand-written OCL queries. Their architecture includes language-specific modules for serializing OCL queries into files and provides built-in mechanisms for storing validation results, including the set of constraint violations and the execution time required for each OCL check.

The AAS is a digital representation of a physical asset. It is composed of multiple submodels that capture the asset's data and functions. It enables various communication interfaces and acts as the bridge between the asset and the digital world (Neidig et al., 2022).

Several studies explore the combination of AAS with semantic constraint mechanisms. For example (Rimaz, 2024) investigates AAS and SHACL-based constraint verification for semantic digital twins, while (Eichelberger & Weber, 2024) addresses model-driven generation of AAS submodels. (Miny et al., 2020) demonstrate how OCL can be applied to generate well-formed AAS submodels, highlighting the applicability of OCL as a formal mechanism for enriching AAS structures with precise semantic constraints.

The *File* submodel element (SME) in the AAS allows the attachment of relevant data files to assets or components (IEC, 2022). File embedding within the AAS has been previously explored; for instance, (Fan et al., 2023) recommend attaching important engineering files to the *File* SME to ensure that all digital information required for planning and executing specific process steps is available directly within the AAS.

AutomationML (AML) allows modelling, storage and exchange of complex technical systems and is designed for exchanging engineering data and covers a multitude of relevant aspects in engineering (IEC, 2018) (Drath, 2021). It is based on comprehensive XML based object-oriented data modeling language. For example, (R. Gudder et al., 2024) propose a data transformation approach which utilizes AML as a standardized data format to transform and integrate engineering data from multiple domains into a single, unified AML data model for a complex production system.

At last, (Alvarado et al., 2025) identify the limitation that the current AAS metamodel does not natively support the representation of OCL rules. Therefore, authors of mentioned article propose mapping OCL validation results into property SMEs of type Boolean, providing a workaround for integrating constraint verification outcomes into the AAS structure.

However, to the best of our knowledge, no existing study investigates how OCL constraints and their corresponding validation outcomes can be managed using AAS. To address this gap, the present article proposes a complete workflow for managing semantic OCL constraints and their validation results within the AAS, by creating the necessary submodels to connect models, and their constraint definitions where AAS is being used as the central integration layer. Hence, authors define both OCL constraints and the corresponding validation results as files and integrate them into dedicated submodels of AAS to ensure interoperability.

## 3. METHODOLOGY

This chapter presents a methodology and conceptual approach for managing semantic OCL constraints and their validation results within the AAS. The goal of the methodology is to enable consistent representation and management of semantic correctness of system's information in a manner that it is machine-interpretable and interoperable across industrial systems like Manufacturing Execution Systems (MES), digital twins or domain systems.

The proposed methodology consists of five conceptual steps (S1, S2, S3, S4, S5). The description and workflow of the steps is shown in Fig. 1:

S1: Semantic representation of system
S2: Constraint representation of system
S3: Integration of semantic and constraint models within AAS
S4: A generic mechanism for constraint validation
S5: Constraint result fetch by MES, digital twins or domain systems

These elements together define how domain knowledge, structural rules, and dynamic runtime information can be jointly represented and evaluated in an industry 4.0 environment.

### 3.1 Description of proposed steps

S1: Semantic representation of system

At the conceptual level, system semantics consists of two parts:

(1) Type-level semantics, which describe the structural and functional characteristics of a system or an asset.

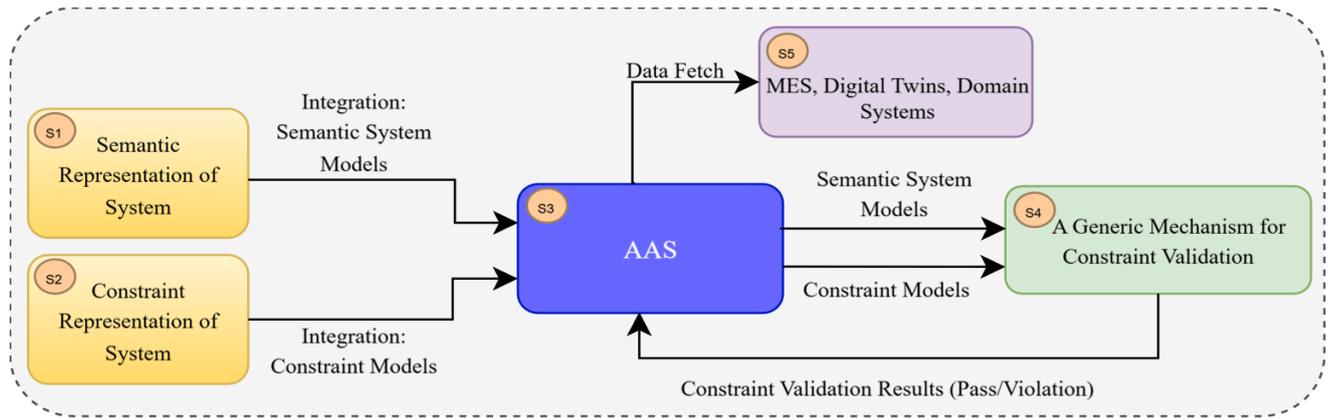

Fig. 1. Conceptual architecture for semantic and constraint integration in the AAS and its 5 conceptual steps S1-S5

(2) Instance-level semantics, which contain the concrete values and configuration information of a specific system or asset.

Both types of semantics form a unified knowledge model. The methodology assumes that these models are expressed in a formal modeling language so that their structure and semantics can be interpreted and validated by software systems.

S2: Constraint representation of system

Semantic models alone cannot capture correctness conditions, dependencies, and logical rules. Therefore, the methodology suggests creating formal semantic constraints that express rules about how system models are expected to behave. These constraints are represented using the formal constraint language (e.g. OCL).

S3: Integration of semantic and constraint models within AAS

The methodology organizes semantics and constraints within the AAS so that other systems can understand and use them. Therefore, the AAS serves as the conceptual container that organizes semantic information models, constraint models and the resulting validation information.

The authors propose the introduction of two dedicated conceptual submodels:

(1) Semantic information submodel: Organizes the type and instance semantics.
(2) Semantic constraint submodel: Organizes the constraint definitions and exposes them in a standardized, machine-readable way.

S4: A generic mechanism for constraint validation

The methodology further defines a workflow that connects semantic modeling, constraints and the AAS in following steps:

(1) A generic validation mechanism retrieves semantic models and constraints from AAS whenever validation is required.
(2) If semantic information models and constraint models are created in different modeling environments, the mechanism should convert the semantic models into a format that the constraint model can interpret for enforcing constraints.
(3) The mechanism evaluates whether the semantic model satisfies the domain constraints.
(4) The validation result (e.g. rule satisfied, violated) is written back into the AAS.

S5: Constraint result fetching by MES, digital twins or domain systems

Once the constraint validation results are generated, other systems such as MES, digital twins, or domain-specific systems can retrieve these results. They use the pass/violation information for tasks such as checking model correctness, ensuring data quality, detecting design or process issues, verifying control logic, speeding up verification processes, and maintaining overall data integrity.

## 4. IMPLEMENTATION WORKFLOW

To support the application of the proposed methodology, a implementation workflow is illustrated in Fig. 2. The conceptual methodology defines the abstract steps (S1-S5), the implementation workflow shows one practical way to execute these steps and demonstrates how AAS structure and validation flow operate. This workflow is a suggestion by authors and can be adapted to individual requirements. The implementation recommends creating two dedicated AASs; one for system's semantic representation and one for constraint representation. A single AAS can also handle both purposes, but separate AASs are proposed here for clarity. The implementation workflow can be carried out as follows:

(1) S1 Type-level and Instance-level semantics can be modeled using AML and integrated in *File* SMEs (*InformationModel_Type, InformationModel_Instance*) of AAS1 (S3).
(2) Dynamic system attributes can be represented into *Property* SMEs (*Dynamic_Attribute1, Dynamic_Attribute2*) within AAS1 (S3).
(3) The *Relationship* SME in S3 can be used to represent the semantic links between dynamic attributes and the corresponding type and instance models.
(4) S2 Constraints can be modeled using OCL and integrated into the *File* SME (*Constraint_Model*) of AAS2 (S3).

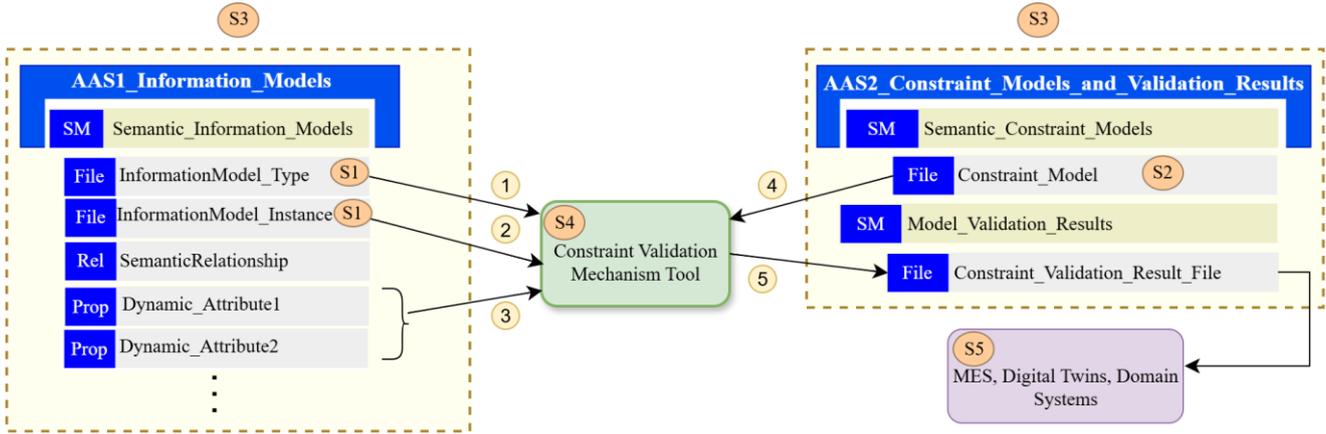

Fig. 2. Workflow for methodology implementation using the proposed AAS Structures and constraint validation mechanism

(5) An OCL Validation Component (OVC) is developed in this study to provide the constraint validation mechanism (S4). The tool is currently under development. In this study, it is specifically designed to process AML files integrated within the AAS.

To validate OCL constraints over AML systems models, the AML models must first be transformed into XML Metadata Interchange (XMI) (Object Management Group (OMG), 2015) format, which is compatible by Eclipse Modeling Framework (EMF). This means type-level AML files are converted into EMF ecore files, and instance-level AML files are converted into xmi files. The tool operates in the following five steps:

1. Fetch type model: Fetches the *InformationModel_Type* file from AAS1 and converts it into ecore format.
2. Fetch instance model: Retrieves the *InformationModel_Instance* file from AAS1 and maps its attribute values into an xmi instance.
3. Inject dynamic values: Retrieves the dynamic attribute values from the *Property* SMEs in AAS1 and updates the corresponding attributes in xmi instance models.
4. Fetch OCL constraints: Retrieves the OCL constraint models from AAS2 and performs OCL validation on xmi instances.
5. Write results: Generates the validation results in text file and serializes them into JSON/XML, storing the output in AAS2 for machine-readable processing.

File fetching, file integration, and dynamic attribute retrieval in the AAS can be performed using REST APIs. For this purpose, Eclipse BaSyx, an open-source platform (Eclipse BaSyx, 2025) is used as it provides a unified interface for deploying AASs and their submodels. A full description of the generic BaSyx AAS HTTP REST APIs can be found in (Eclipse BaSyx Project, 2020).

## 5. APPLICATION SCENARIO

To demonstrate the applicability of proposed methodology, a small fictional industrial scenario is implemented. The scenario represents a simplified production facility consisting of four sequential manufacturing processes. The purpose of the implementation is to show semantic information models and formal OCL constraints can be represented, validated, and managed using AAS. As stated in (Industrial Digital Twin Association (IDTA), 2020), an industrial facility can be considered an asset with its own AAS and can be accessed through its ID. This allows an entire system to be modeled in the AAS by representing it as a single asset or component.

All files created for the demonstration are available in a public GitHub repository[1].

### 5.1 Industrial use case description

The production system consists of four intermediate processes, as shown in Table 1, which are executed on the shop floor to manufacture the final product. In the setup, three constraints have been considered, as shown in Table 2. Among the mentioned constraints, the *AppropriateTemperature* constraint has already been solved in terms of control technology. Here this constraint is just taken for example purpose.

The remaining two constraints are newly introduced and play a critical role in industrial scenarios, particularly in process rescheduling. These constraints are highly relevant because process rescheduling is often required in manufacturing settings for several purposes, for example, optimizing energy costs by shifting energy-intensive processes to periods of lower energy prices, or managing machine breakdowns, where all tasks assigned to a failed machine must be reassigned to alternative time slots. A detailed description of these files, along with additional relevant information, is provided in Table 3. All created semantic and constraint models of the system, along with their validation results, are integrated into AAS instances created specifically for this study. Table 4 shows this integration.

Table 1. Manufacturing processes in the production scenario with their attributes and values

| Process | Process Type | Process Sequence Order |
|---|---|---|
| IntermediateProcessA | A | 1 |
| IntermediateProcessB | B | 2 |
| IntermediateProcessC | C | 3 |
| IntermediateProcessD | D | 4 |

Table 2. Constraints name, description and their importance in manufacturing facility

| Constraint | Constraint Name | Constraint Description | Constraint importance |
|---|---|---|---|
| 1. | UniqueProcessOrderConstraint | Ensures that the sequence order value of the manufacturing processes is always unique | This ensures the proper execution of ProcessSequenceConstraint |
| 2. | ProcessSequenceConstraint | Ensures that the execution of manufacturing processes remains fixed and cannot be altered | The dependencies between the manufacturing process steps are necessary so that the MES knows which sequence to follow in case of process rescheduling. |
| 3. | AppropriateTemperature | Ensures that the temperature of manufacturing facility should be appropriate | An appropriate shop floor temperature is required to avoid defects in final product. |

Table 3. Created model files and generated constraint result files with designated filenames

| Aspect/Purpose | Semantic Representation of System (AML) | Constraint Representation of System (EMF/OCL) |
|---|---|---|
| Type-level Model | *DemoProductionProcessesTypemodel.aml* — defines generic process structure and shopfloor hierarchy | *demo.ecore* — defines type-level system structure + OCL constraints (constraint logic) |
| Instance-level Model | *DemoShopfloorInstanceModel.aml* — concrete values for processes and facility | *Demo_ShopFloor_Successful*.xmi and *Demo_ShopFloor_Violated*.xmi — instance values used for validation |
| Static Attributes | Defined directly in AML instance model (e.g., *processType*, *processSequenceOrder*) | Present in XMI instance models; validated by OCL constraints |
| Dynamic Attributes | Represented using AAS *Property* SME (e.g., *currentTemperature*), which can be updated in real time | Not stored in XMI; OCL can reference them through integration |
| Validation Artifacts | ——— | Validation result files (text + JSON), serialized via *ResultSerialization.py* |

Fig. 3 depicts the structure of the AAS created to represent semantic system model, while Fig. 4 shows the AAS structure developed to represent the constraint models and corresponding validation results. This setup demonstrates how an AAS can serve as a unified container for both semantic model info-formation and validation data.

### 5.2 Mechanism for constraint validation

In this example, the transformations between AML, Ecore, XMI as well as the constraint evaluations, are performed manually. In future, the authors aim for automatic execution of these steps by means of the prototype OVC tool, enabling real-time semantic validation directly based on the AML models integrated in AAS.

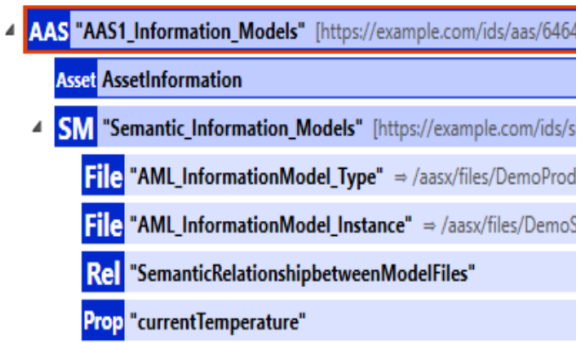

Fig. 3. Information models AAS, showing submodel structure and integrated files.

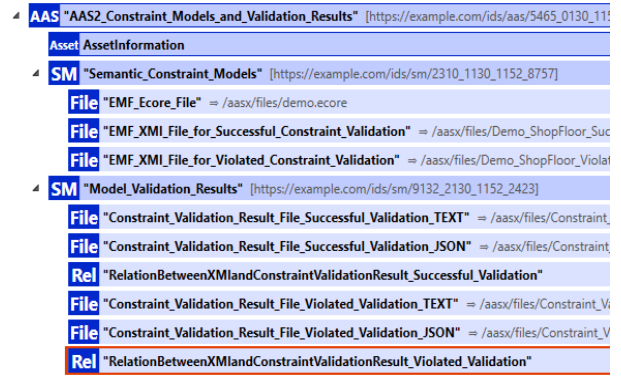

Fig. 4. Constraint models and validation results AAS, showing submodels structure and integrated files.

Table 4. AAS submodel structures for system and constraint representation and their integrated files for use

| AAS Instance | Submodel | Contents/Purpose |
|---|---|---|
| AAS1-Information Models | Semantic Information Models | • AML_InformationModel_Type (type-level semantics)<br>• AML_InformationModel_Instance (instance-level semantics)<br>• Dynamic attribute (*currentTemperature*) in property SME<br>• Semantic relationships between model files (type and instance) |
| AAS2-Constraint Models and Validation Results | Semantic Constraint Models | • Ecore model (type-level strucutre + OCL constraints)<br>• XMI instance models (created for both, successful and violated constraint validation) |
|  | Model Validation Results | • Validation result files (text + Json; text file is serialized into Json format using script *ResultSerialization.py*)<br>• *Relationship* SMEs linking xmi instance models with their corresponding validation outcome files. |

## 5.3 Result and Discussion

The validation result files (text and JSON) integrated in the submodel (*Model_Validation_Results*) as in Fig. 4, show the results of applying the constraints to the two instance XMI files stored in the *File* SME of the submodel (*Semantic_Constraint_Models*) in the same AAS. In the successful instance, all constraints were validated without any violations, confirming that the modeled process sequence and temperature values are correct. In contrast, the violating instance triggered errors for all three constraints, showing non unique process orders, an incorrect sequence, and a temperature above the allowed limit.

These results show that OCL constraints can be reliably managed using the AAS, which supports automated quality checks in MBSE environments. The article also highlights the potential of the OVC to reduce manual effort and improve automation in constraint validation.

## 6. SUMMARY AND OUTLOOK

This paper presents a method and a workflow for managing semantic OCL constraints and their validation results using the AAS as a central integration layer in the MBSE context. The workflow was demonstrated through a fictional industrial use case. All models and files created for this example are available in a public GitHub repository[1]. An automatic constraint validation tool, the OVC, was introduced to support model transformation, constraint execution, and result storage, reducing manual effort.

Future work aims to further develop the OVC tool to enable fully automated OCL constraint validation and result extraction, as well as extending the set of constraints to cover real industrial scenarios.

## ACKNOWLEDGEMENTS

This work was carried out as part of the greenProd project funded by the Federal Ministry for Economic Affairs and Climate Protection under the funding code '01MN23003B', and the authors would like to express their thanks for the funding. The responsibility for the content of this work lies with the authors. OpenAI's ChatGPT was used to check grammar and spelling.